\documentclass{mn2e} \usepackage{psfig}

\def\etal{et~al. }
\def\la{\mathrel{\hbox{\rlap{\hbox{\lower4pt\hbox{$\sim$}}}{\raise2pt\hbox{$<$}}}}}
\def\ga{\mathrel{\hbox{\rlap{\hbox{\lower4pt\hbox{$\sim$}}}{\raise2pt\hbox{$>$}}}}}

\begin{document}

\title[Rotational period of WD1953-011] {Rotational period of
WD1953-011 - a magnetic white dwarf with a star spot}

\author[C.\,S. Brinkworth \etal] {C.\,S. Brinkworth$^1$,
T.\,R. Marsh$^{2}$, L. Morales-Rueda$^3$, P.\,F.\,L. Maxted$^4$,
\newauthor M.\,R. Burleigh$^5$, S. \,A. Good$^5$\\ \\ $^1$ Department
of Physics and Astronomy, University of Southampton, Highfield,
Southampton, SO17 1BJ, UK. \\ $^2$ Department of Physics, The
University of Warwick, Coventry, CV4 7AL, UK. \\ $^3$Department of
Astrophysics, University of Nijmegen, PO Box 9010,  6500GL, Nijmegen,
The Netherlands \\ $^4$ School of Chemistry and Physics, Keele
University, Staffordshire, ST5 5BG, UK \\ $^5$ Department of Physics
and Astronomy, University of Leicester, Leicester, LE1 7RH, UK. \\}


\maketitle

\begin{abstract}
WD1953-011 is an isolated, cool ($7920\pm 200\mathrm{K}$, Bergeron,
 Legget $\&$ Ruiz, 2001) magnetic white dwarf (MWD) with a low average
 field strength ($\sim 70\mathrm{kG}$, Maxted \etal 2000) and a higher
 than average mass ($\sim 0.74\mathrm{M}_{\odot}$, Bergeron \etal
 2001). Spectroscopic observations taken by Maxted \etal 2000 showed
 variations of equivalent width in the Balmer lines, unusual in a low
 field white dwarf. Here we present V band photometry of WD1953-011
 taken at 7 epochs over a total of 22 months. All of the datasets show
 a sinusoidal variation of approximately 2\% peak-to-peak
 amplitude. We propose that these variations are due to a star spot on
 the MWD, analogous to a sunspot, which is affecting the temperature
 at the surface, and therefore its photometric magnitude. The
 variations have a best-fit period over the entire 22 months of 1.4418
 days, which we interpret as the rotational period of the WD.

\end{abstract}


\section{Introduction}

There are over 120 catalogued isolated magnetic white dwarfs (MWD),
comprising $\sim$ 2$\%$ of the total WD population. Their field
strengths range from 10kG up to 1000MG (Wickramasinghe \& Ferrario,
2000), with temperatures ranging from $\sim$4000K to $>$50000K. MWDs
are important from an evolutionary point of view as they tend to have
a higher than average mass than their non-magnetic counterparts,
suggesting that the magnetic field affects the initial-to-final mass
relationship. They are also extremely useful for determining spin
periods, as rotation in non-magnetic white dwarfs is notoriously hard
to measure due to the heavy gravitational broadening of their spectral
lines. In contrast, a significant fraction of magnetic white dwarfs
($\sim30\%$) display spectroscopic, spectropolarimetric and/or
photometric variability indicative of rotation.  Spectral or
spectropolarimetric variation is generally believed to be caused by
surface field strength variation (e.g. motion of Zeeman-split
components of the H Balmer absorption lines), while photometric
variability in high-field MWDs is due to the field dependence of the
continuum opacity (magnetic dichroism, Ferrario \etal 1997). Low-field
MWDs are not expected to show significant photometric variability.

The measured rotational periods of MWDs are unusual as they seem to
show a bimodal distribution, with one group rotating very slowly,
possibly with periods $>100$years, and another group rotating very
quickly, of order minutes to hours. This contrasts with the estimates
of the rotational periods of non-magnetic WDs, which suggest
timescales of $\sim$1 day (Heber, Napiwotski $\&$ Reid, 1997; O'Brien
\etal 1996). These results suggest efficient angular momentum transfer
from the core to the envelope and large-scale angular momentum loss
during post main-sequence evolution, otherwise WDs should be rotating
close to their break-up value. For MWDs, Spruit (1998) proposed that
the extremely slow rotators could be produced if the magnetic field
locks the forming MWD to its envelope, efficiently shedding angular
momentum, while King, Pringle $\&$ Wickramasinge (2001) have suggested
that the very fast rotators ($P_{rot}\sim$ minutes) may have been spun
up in double-degenerate mergers.

WD1953-011 is a cool (7920$\pm$200K, Bergeron \etal 2001) magnetic
white dwarf with a low field strength ($\sim$70kG). The field
structure of MWDs can usually be modelled with a centred or offset
dipole, but Maxted \etal (2000) have shown the field structure of
WD1953-011 to be more complex, with a strong ($\sim$500kG) spot-like
field superimposed on a weaker ($\sim$70kG) dipolar
distribution. Maxted \etal (2000) also discovered changes in the
equivalent width of the Balmer lines with time. Here we show that
WD1953-011 is also photometrically variable, and we use those
variations to find the rotational period of the MWD.

\section{Observations}
We observed WD1953-011 at 7 epochs between July 2001 and May 2003. In
total we obtained 900 observations in the V band. The data were all
taken using the 1m Jacobus Kapteyn Telescope on La Palma. A full list
of V band observations is given in Table 1, although we also took data
in B, R and I broad bands and H$\beta$ narrow band during the July
2001 run. As the results of the photometry were found to be almost
identical in all of the bands, we restricted the subsequent
observations to V band only. The SITe1 CCD chip is 2088 x 2120 pixels,
with readout noise = 6 e and gain = 1.9 e/ADU. Pixel size is 15$\mu m$
and image scale is 0.33''/pix. Fig 1 shows a finding chart for WD1953-011
and the comparison stars.

\begin{figure}
\begin{picture}(10,0)(10,10)
\put(0,0){\includegraphics{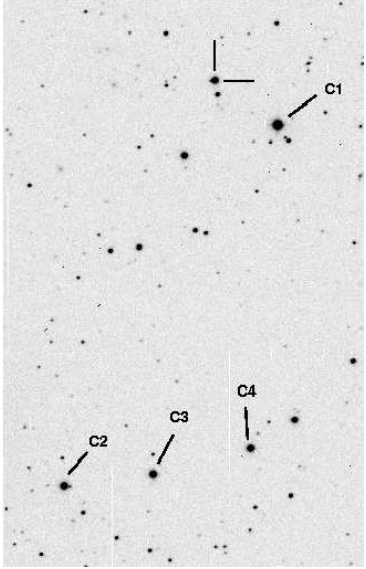}}
\noindent
\end{picture}
\vspace{80mm}
\caption{Finding chart for WD1953-011 and the 4 comparison stars used.}
\label{fig1}
\end{figure}

\begin{table}
\caption{List of observations of WD1953-011 taken with the JKT on La
Palma. Observers: C S Brinkworth CSB, T R Marsh TRM, L Morales-Rueda
LMR, M R Burleigh MRB, S A Good SAG}

\begin{tabular}{cccccc}
Dates & Filter & Exp (s) & N & Observer & Conditions \\   & & & & & \\
05/07/01 & V Kitt & 40 & 14 & TRM & Good\\  06/07/01 & V Kitt & 40 &
18 & TRM & Fair \\   07/07/01 & V Kitt & 40 & 19 & TRM & Superb \\
08/07/01 & V Kitt & 40 & 6 & TRM & Poor seeing \\   09/07/01 & V Kitt
& 40 & 9 & TRM & Good \\   10/07/01 & V Kitt & 40 & 12 & TRM & Good \\
11/07/01 & V Kitt & 40 & 12 & TRM & Good \\   14/05/02 & V Harris & 40
& 30 & MRB,SAG & Cirrus \\   15/05/02 & V Harris & 40 & 45 & MRB,SAG &
Good \\   16/05/02 & V Harris & 40 & 25 & MRB,SAG & Variable \\
17/05/02 & V Harris & 40 & 40 & MRB,SAG & Good \\   26/05/02 & V Harris
& 60 & 15 & TRM,CSB & Good \\   27/05/02 & V Harris & 60 & 10 &
TRM,CSB & Good \\ 28/05/02 & V Harris & 60 & 20 & TRM,CSB & Cirrus \\
29/05/02 & V Harris & 60 & 20 & TRM,CSB & Cirrus \\   30/05/02 & V
Harris & 60 & 30 & TRM,CSB & Cirrus \\   31/05/02 & V Harris & 60 & 20
& TRM,CSB & Good \\  01/06/02 & V Harris & 60 & 20 & TRM,CSB & Good \\
15/07/02 & V Harris & 60 & 26 & LMR & Good \\   16/07/02 & V Harris &
60 & 36 & LMR & Good \\   17/07/02 & V Harris & 60 & 11 & LMR & Good
\\ 18/07/02 & V Harris & 60 & 25 & LMR & Good \\   19/07/02 & V Harris
& 60 & 27 & LMR & Good \\   20/07/02 & V Harris & 60 & 29 & LMR & Good
\\ 21/07/02 & V Harris & 60 & 34 & LMR & Good \\   02/08/02 & V Harris
& 60& 5 & MRB,CSB & Good \\   03/08/02 & V Harris & 60 & 20 & MRB,CSB
& Good \\   04/08/02 & V Harris & 60 & 15 & MRB,CSB & Good \\ 05/08/02
& V Harris & 120 & 15 & MRB,CSB & Some cirrus \\   06/08/02 & V Harris
& 40 & 15 & MRB,CSB & Superb \\   07/08/02 & V Harris & 60 & 5 &
MRB,CSB & Good \\   10/09/02 & V Harris & 60 & 30 & TRM & Good \\
11/09/02 & V Harris & 60 & 24 & TRM & Cirrus \\   12/09/02 & V Harris
& 60 & 32 & TRM & Superb \\   13/09/02 & V Harris & 60 & 40 & TRM &
Some cirrus \\   15/09/02 & V Harris & 100 & 20 & TRM & Poor seeing \\
08/05/03 & V Harris & 60 & 3 & LMR & High cloud \\ 09/05/03 & V Harris
& 60 & 4 & LMR & Good \\ 10/05/03 & V Harris & 60 & 5 & LMR & Good \\
11/05/03 & V Harris & 60 & 5 & LMR & Poor seeing \\ 12/05/03 & V
Harris & 60 & 5 & LMR & Good \\ 13/05/03 & V Harris & 60 & 5 & LMR &
Good \\ 14/05/03 & V Harris & 60 & 5 & LMR & Good \\ 15/05/03 & V
Harris & 60 & 5 & LMR & Good \\ 16/05/03 & V Harris & 60 & 10 & LMR &
Twilight \\ 17/05/03 & V Harris & 60 & 8 & LMR & Twilight \\ 18/05/03
& V Harris & 60 & 5 & LMR & Variable seeing \\ 20/05/03 & V Harris &
60 & 6 & LMR & High cloud \\ 21/05/03 & V Harris & 60 & 5 & LMR & Good
\\ 22/05/03 & V Harris & 60 & 5 & LMR & Dusty \\ 23/05/03 & V Harris &
60 & 5 & LMR & Good \\ 25/05/03 & V Harris & 60 & 5 & CSB, MRB & Good
\\ 26/05/03 & V Harris & 90 & 5 & CSB, MRB & Superb \\ 27/05/03 & V
Harris & 60 & 5 & CSB, MRB & Superb \\ 28/05/03 & V Harris & 60 & 10 &
CSB, MRB & Clear but dusty \\ 29/05/03 & V Harris & 60 & 5 & CSB, MRB
& Good \\ 30/05/03 & V Harris & 60 & 5 & CSB, MRB & Good \\ 31/05/03 &
V Harris & 60 & 10 & CSB, MRB & Superb \\

\end{tabular}
\end{table}

\begin{figure*}
\begin{picture}(470,0)(10,20)
\put(0,0){\includegraphics{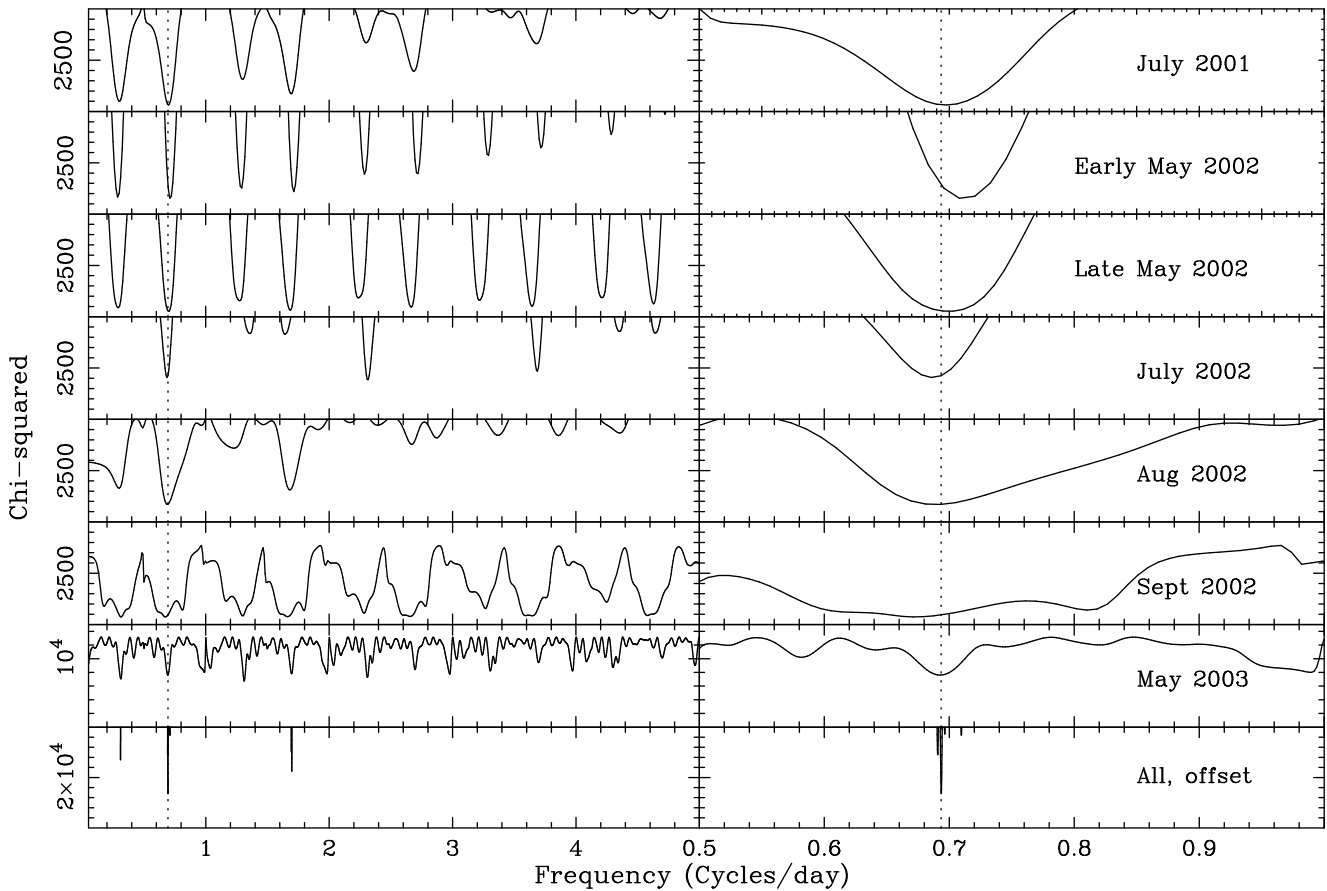}}
\noindent
\end{picture}
\vspace{125mm}
\caption{Periodograms for all of the 7 data sets. A period of
  approximately 1.4418 days is favoured (vertical dotted line). The
  expanded plots seem to show a slight shift in the best-fitting
  period (see Fig 4 and Section 4.2).}
\label{fig2}
\end{figure*}

\section{Data Reduction}
Each of the seven data sets were reduced in the same way using the
packages FIGARO and KAPPA. First the bias frames from a run were
combined to form a master bias for the whole run. This was subtracted
from all other frames. The flats were then checked, and those with
mean counts of less than 7000, or greater than 35000 were
discarded. We were concerned that there may be a problem with the flat
fields at short exposure times, caused by the shutter speed allowing
the centre of the chip to be exposed for a longer time than the
edges. This was potentially important as the variation we were trying
to measure between the target and comparisons was small - of order
2$\%$. The flats were checked by dividing those of different exposure
time by each other, and we found a gradient peaking in the centre of
the chip. However, this effect was only seen in the September 2002
data, and was not only confined to the very low exposure times
($<2\mathrm{s}$) as we were expecting, but affected flats with
exposures of up to 20s. We suspect that this was caused by a loose
filter moving with respect to the chip, but that these variations
smoothed out over long exposure times. We therefore only used flats
with exposure times of $> 20\mathrm{s}$ to generate the master flat.

\begin{figure}
\begin{picture}(10,0)(10,10)
\put(0,0){\includegraphics{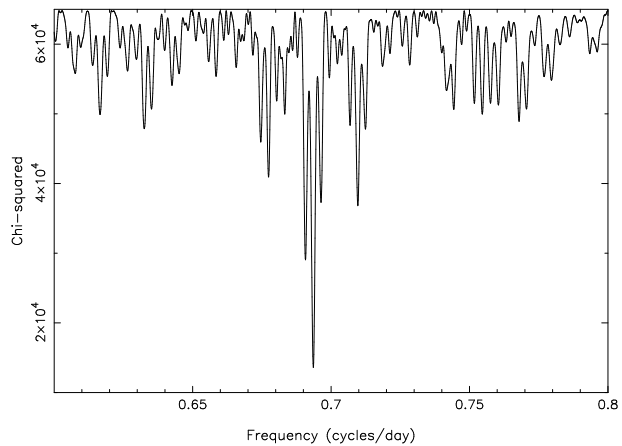}}
\noindent
\end{picture}
\vspace{60mm}
\caption{Periodogram zoomed in on the best-fitting period of 1.44176
  days (0.6935 cycles/day).}
\label{fig3}
\end{figure}

A single master flat was generated for each run in an attempt to
remove systematic variations from night to night. This was not
possible for the July 2001 run, as the first night's data was taken on
a different part of the chip to the rest of the run, and there were no
full-frame flats. Therefore the first night of that run is flatfielded
with a different master flat to the other 6 nights. The first May 2002
run was flat fielded with dome flats as there were no sky flats
available. All of the dome flats had exposure times of 10 seconds. The
second May 2002 run had target frames with a mixture of fast and slow
readout speeds. We therefore used two different master flats, one for
the fast readout frames and one for the slow. The May 2003 run was
during a very dusty period, hence the flats changed nightly. This run
was therefore flatfielded with an individual master flat for every
night.

Once the master flats had been divided from the target frames, we
performed aperture photometry using AUTOPHOTOM. This was performed
with several different apertures to determine the optimum aperture
radius of 4 pixels. The sky background was taken from an annulus
around the target stars, the measurement errors were estimated from
sky variance and the sky background level was estimated using the
clipped mean of the pixel values in the annulus.

Results were output in counts. Once we had established that the three
comparison stars were not varying, we combined their fluxes to give us
one bright comparison star, and divided the target photometry by the
newly generated comparison star to give us differential photometry of
the target.

The July 2001 data had been taken with the Kitt Peak V filter, while
the rest of the data was taken with the Harris V filter. We corrected
for this by integrating models for a cool WD (for the target) and a
G-type MS star (for the comparisons) through both filter responses,
and multiplying the July 2001 data by the ratio. The correction to the
differential photometry only amounted to a factor of 0.9974.

All times were corrected to heliocentric Julian days.

\section{Analysis and Results}

\subsection{Determining the periods}

We used a ``floating mean'' periodogram (e.g. Cumming, Marcy $\&$
Butler, 1999; Morales-Rueda \etal 2003) to determine the period of
each epoch separately, and all of the data together. This is a
generalisation of the Lomb-Scargle periodogram (Lomb, 1976; Scargle,
1982) and involves fitting the data with a sinusoid plus constant of
the form:

\[
 A + B \sin[2\pi f(t - t_{0})],
\]

\noindent
where $f$ is the frequency and $t$ is the observation time. The
advantage over the Lomb-Scargle periodogram is that it treats the
constant, A, as an extra free parameter rather than fixing the
zero-point and then fitting a sinusoid, i.e. it allows the zero-point
to ``float'' during the fit. The resultant periodogram is an inverted
$\chi^2$ plot of the fit at each frequency (Figs 2 \& 3).

\subsection{Uncertanties in the periods}

The errors output by the packages are the formal statistical errors,
but due to the high signal to noise they are likely to be
underestimates of the actual errors due to e.g. anomalies in the flat
fields, aperture edge effects, and so on. We therefore obtained an
independent estimate of our errors by bootstrapping our data. We fit
the data for each epoch with a sine wave, then resampled the data,
randomly selecting the same number of points and re-fitting with the
sine wave (Diaconis \& Efron, 1983). This was repeated 500000 times. The
resultant period distributions can be seen in Fig 5. In order to avoid
excessive weighting of a few data points, the errors were set to a
standard average value before bootstrapping. The bootstrapping seems
to indicate that there is a small change in the best-fitting period
between each epoch. To test the robustness of this result to
night-to-night systematic shifts we repeated the bootstrap runs after
adding offsets to each night.  The offsets were added as Gaussian
random variables. We found that an RMS offset of only 0.003 magnitudes
caused enough of a spread in the period distributions that the period
shift between each epoch was no longer significant. As such a shift
could be caused by anomalies in the flat fields, irregularities in the
chip or by slight variations in the standard stars, we conclude that
there is no evidence for a period change in WD1953-011 in our data.

The phase-folded light curve (Fig 4) shows a variation in the flux of
$\pm1\%$. Fig 2 shows the periodograms for each epoch, showing that
the deepest minimum in $\chi^2$ for all but one data set, and the only
minimum common to all epochs, is that at approximately 0.69 cycles per
day, corresponding to:
\[
HJD = 2452489.3588(9) + 1.441769(8)E
\]
which specifies the time of minimum light.  The zero-point was
selected to give the minimum correlation between it and the fitted
period. The light curve also appears to be slightly non-sinusoidal at
the level of 1-2 mmag in the 1st harmonic. There initially appear to be several
sharp features in the folded light curve, most notably at phase
0.4. However, all of these outlying points are from single nights during either the first
May 2002 or the July 2002 run. As these features are not seen at any
of the other epochs, we believe that they are not significant
features in the WD1953-011 light curve.

\begin{figure}
\begin{picture}(10,0)(10,10)
\put(0,0){\includegraphics{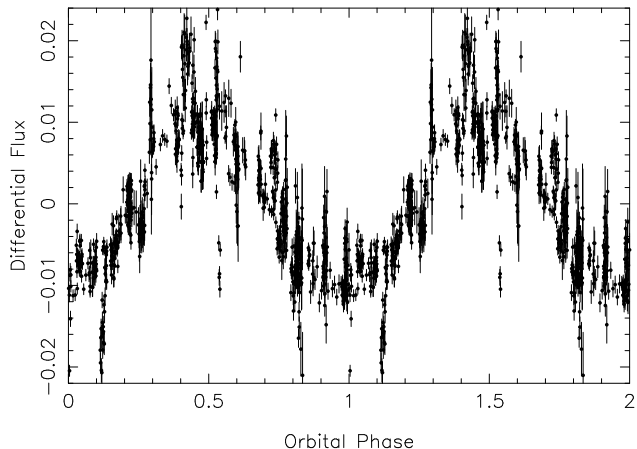}}
\noindent
\end{picture}
\vspace{60mm}
\caption{All of the data folded on the best-fitting period of 1.44176
days.}
\label{fig4}
\end{figure}

\begin{figure}
\begin{picture}(10,0)(10,10)
\put(0,0){\includegraphics{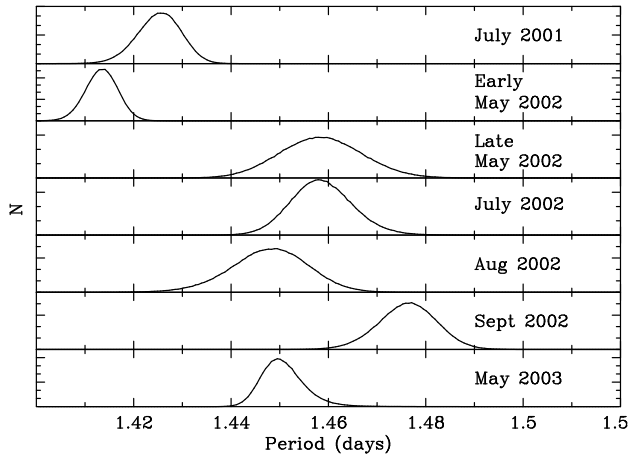}}
\noindent
\end{picture}
\vspace{58mm}
\caption{Period distributions for all 7 data sets after bootstrapping
  500000 times and plotting over 200 bins (see Section 4.2).}
\label{fig5}
\end{figure}

\section{Discussion and Conclusions}

The variation in the flux from WD1953-011 could be explained if it
were a binary system, with the secondary emitting re-processed light
visible for part of the orbital cycle. However, radial velocity
measurements by Maxted \etal (2000) found that it was stable to within
2 km/s. Within this error, and using an orbital period of 1.44 days,
it is still possible to miss a companion body with mass $<
0.009\mathrm{M}_{\odot}$ ($\sim$10M$_{J}$) orbiting at approximately
0.02 AU. However, as the WD is so cool with a relatively low UV flux,
such a body would only re-process $\sim0.013\%$ of the light from the
white dwarf, and hence could not produce the variability on the
$\sim2\%$ peak-to-peak level that we see. This  leads us to believe
that the variability is somehow caused by the magnetic field of the
MWD.  Periodic photometric variation has been seen before in white
dwarfs with a high magnetic field strength, such as RE J0317-853
(Barstow \etal 1995). In these cases the variation is thought to be
due to the field dependence of the continuum opacity (magnetic
dichroism, Ferrario \etal 1997), but the field strength of WD1953-011
($\sim$70kG) is not large enough to cause this effect. Instead we
believe that the variations may be caused by a star spot on the
surface of the WD, analogous to a sun spot. Star spots occur when the
atmospheric convection of the stellar atmosphere is inhibited by the
magnetic field at the surface, so causing a spot of lower temperature,
and therefore lower luminosity, to be formed. As the MWD rotates, the
visibility of this cooler spot will vary, causing a periodic variation
in the flux from the star. At a temperature of only $\sim$7900K,
WD1953-011 is well below the limit required for a convective
atmosphere (15000K, Bergeron, Wesemael \& Beauchamp, 1995), and
therefore may be capable of forming star spots.

The visibility of the spot will depend upon the angle between the spin
axis and our line of sight. Consider a small spot at latitude $\beta$
on the white dwarf. The variation in the light curve will mainly
depend upon the varying projected area of the spot as the white dwarf
rotates (we ignore limb darkening as a second order effect).

Defining phase $\phi$=0 as the point at which spot is closest to us,
then the cosine of the angle between the normal to the surface of the white dwarf at the
location of the spot and our line of sight ($\alpha$) is given by

\[ \cos\alpha = \cos \beta \sin i \cos \phi + \cos i \sin \beta,\]
where $i$ is the inclination of the spin axis to our line of sight. The projected area factor, $\cos
\alpha$, therefore varies sinusoidally with the phase $\phi$. This will be true so long as cos
$\alpha$ $>$ 0 for all $\phi$ (if $\cos$ $\alpha$ $<$ 0 then the spot is
not visible, so the light curve will be flat).

Therefore, for a sinusoidally varying light curve, we require that
$\beta$ $> i$. If seen at large $i$ then the spot
must be near the pole, but if $i$ is small, then the spot could be
almost anywhere on the visible hemisphere - the only condition is that $\beta$
$> i$.

The amplitude of the light curve depends upon the size of the spot and how dark it is, and it would be easy to fit the light curve for a variety of spot sizes. Since a spot of finite size is simply
the result of integrating many infinitesimal spots, large spots can also
lead to sinusoidal variations as long as every part of them satisfies the
$\beta > i$ constraint. This is consistent with the
model proposed by Maxted \etal (2000), who suggested that the magnetic
spot may cover $\sim10$\% of the surface of the WD. Limb-darkening of the form $I \propto 1 - \epsilon + \epsilon \cos
\alpha$ will introduce a first harmonic from the $\epsilon \cos
\alpha$ factor. This will be negligible as long as
\[ \frac{1}{2} \epsilon \cos \beta \sin i \ll 1 - \epsilon,\]
where $\epsilon$ is the linear limb-darkening coefficient. Taking
$\epsilon \approx 0.6$, we require $\cos \beta \sin i \ll 1.3$.
This can be satisfied along with $\beta > i$ by many values of spin
axis inclination and spot latitude, e.g. $i = 30$, $\beta = 70$ gives
$\cos \beta \sin i = 0.17$. In this case, a large spot would help
suppress the harmonic term relative to the fundamental. Thus a spot on
the surface provides a natural explanation for the sinusoidal
flux variation that we see.

It has been suggested that the observed variations may be caused
instead by the presence of circumstellar matter caught in the magnetic
field of the WD as observed in some helium-rich Bp stars (e.g. Groote
\& Hunger, 1982). We believe that this is highly unlikely due to the
absence of emission lines in the spectra of the star taken by Maxted
\etal (2000), and the absence of a formation mechanism for these
clouds. Three possible origins are suggested in Groote \& Hunger
(1982): that the clouds are left over matter from the formation of the
WD; that they are formed through accreted matter; or that they are
formed from mass lost by the WD. The first scenario is unlikely as any
matter left over from the formation of the WD should have been driven
off by radiation pressure while the WD was still very hot. Similarly,
the second mechanism should produce emission lines in the MWD spectrum
that are not seen in the observed spectra. Finally, the low
temperature and low magnetic field strength of WD1953-011 make it
doubtful that the stellar wind would be strong enough to drive mass
loss from the WD, or that the ejected mass would be trapped by the
field lines. We therefore find it improbable that the variations seen
in WD1953-011 are caused by anything other than a feature on the
surface of the WD itself.

\section{Acknowledgements}
CSB and SG acknowledge the support of PPARC studentships. TRM, LMR and
MRB also acknowledge the support of PPARC. The Jacobus Kapteyn
Telescope  is operated on the island of La Palma by the Isaac Newton
Group in the Spanish Observatorio del Roque de los Muchachos of the
Instituto de Astrofisica de Canarias. This research has made use of
the SIMBAD database, operated at CDS, Strasbourg, France.


\begin{thebibliography}{99}

\bibitem{} Barstow M.A., Jordan S., O'Donoghue D., Burleigh M.R.,
Napiwotski R., Harrop-Allin M.K., 1995, MNRAS, 277, 971

\bibitem{} Bergeron P., Leggett S.K., Ruiz M.T., 2001, ApJS, 133, 413

\bibitem{} Bergeron P., Wesemael F. Beauchamp A., 1995, PASP, 107, 1047

\bibitem{} Cumming A., Marcy G.W., Butler R.P., 1999, ApJ, 526, 890

\bibitem{} Diaconis P., Efron B., 1983, Sci. Am., 248, No. 6, 96 

\bibitem{} Ferrario L., Vennes S., Wickramasinghe D.T., Bailey J.,
Christian D.J., 1997, MNRAS, 292, 205

\bibitem{} Groote, D., \& Hunger, K., 1982, A\&A, 116, 64

\bibitem{} Heber U., Napiwotski R., Reid I.N., 1997, A\&A ,323, 819

\bibitem{} King A.R., Pringle J.E., Wickramasinghe D.T., 2001, MNRAS,
320, L45

\bibitem{} Maxted P.F.L., Ferrario L., Marsh T.R., Wickramasinghe
D.T., 2000, MNRAS, 315, L41

\bibitem{} Lomb N.R., 1976, Ap\&SS, 39, 447

\bibitem{} Morales-Rueda L., Maxted P.F.L., Marsh T.R., North R.C.,
Heber U., 2003, MNRAS, 338, 752

\bibitem{} O'Brien M.S., Clemens J.C., Kawaler S.D., Dehner B.T.,
1996, ApJ, 467, 397

\bibitem{} Scargle J.D., 1982, ApJ, 263, 835

\bibitem{} Spruit H.C., 1998, A\&A, 333, 603

\bibitem{} Wickramasinghe D.T., Ferrario L., 2000, PASP, 112, 873


\end{thebibliography}
\end{document}